\documentstyle[aps,epsfig,multicol]{revtex} 

\newcommand{\be}{\begin{equation}}
\newcommand{\ee}{\end{equation}}
\newcommand{\bea}{\begin{eqnarray}}
\newcommand{\eea}{\end{eqnarray}}
\newcommand{\lb}{\left[}
\newcommand{\rb}{\right]}
\newcommand{\lp}{\left(}
\newcommand{\rp}{\right)}

\newcommand{\p}{\partial}
\newcommand{\rd}{\mbox{d}}

\begin{document}

\twocolumn[\hsize\textwidth\columnwidth\hsize\csname @twocolumnfalse\endcsname
\title{Quantized Adiabatic Charge Transport in a Carbon Nanotube}
\author{V. I. Talyanskii, Dmitry Novikov$^a$, B. D. Simons, 
L. S. Levitov$^a$}
\address{Cavendish Laboratory, Madingley Road, Cambridge CB3\ OHE, U.K.\\
{\it (a)} Department of Physics, Center for Materials Sciences \& 
Engineering,\\
Massachusetts Institute of Technology, Cambridge, MA 02139}

\maketitle

\begin{abstract}
The coupling of a metallic Carbon nanotube to a surface acoustic wave (SAW)
is proposed as a vehicle to realize quantized adiabatic charge transport
in a Luttinger liquid system. We 
demonstrate that electron backscattering by a periodic SAW potential,
which results in  miniband formation, 
can be achieved at energies near the Fermi level. 
Electron interaction, treated in a Luttinger liquid framework, 
is shown to enhance minigaps and thereby improve current quantization.
Quantized SAW induced 
current, as a function of electron density, changes sign at half-filling.
\vskip2mm
\end{abstract}
]
\bigskip 

\narrowtext


The mechanism of quantized adiabatic transport, as first conceived by 
Thouless~\cite{Thouless}, involves a one-dimensional (1D) electron system 
in a periodic potential that, via backscattering, opens a gap in the 
electron spectrum. If the potential varies slowly and periodically in time in 
such a way that the Fermi level lies within a gap of the instantaneous 
Hamiltonian, then an integer charge $me$ is transported across the system 
during a single period. This results in a quantized current $j=mef$, where 
$f$ is the frequency of the external field. 
If realized 
experimentally, such a device would present an important application as a 
current standard. 

Electron properties of real 1D conductors, such as nanotubes or quantum wires,
are dominated by electron interactions\cite{NT-luttinger,KaneBalentsFisher,Bockrath'99,Yao'99}. 
However, apart from general statements~\cite{ThoulessNiu} about robustness
of the quantization, the effect of interactions on quantized 
transport has not been explored. In this article we establish the possibility
to realize this regime in metallic nanotubes, 
the purest 1D conductors\cite{Tans,Bockrath'97,McEuen'99}
currently available.
We develop a theory that takes full account of 
electron interactions in this system in the Luttinger liquid framework.




Although the quantized adiabatic transport mechanism is compelling in its 
simplicity, is has proven difficult to realize experimentally: the goal is
to find such a combination of a host $1D$ system and a sliding external 
perturbation to engineer a miniband spectrum with minigaps sufficiently 
large that disorder, thermal excitations, and finite size effects do not 
compromise the integrity of the quantization. Recently, a surface acoustic 
wave (SAW) was used to achieve quantized current in a split gate point 
contact~\cite{Talyanskii'96}. 
The SAW field can be strong enough to induce a bulk gap, and the 
SAW wavenumber can be chosen to match $2 p_F$ to pin electrons. Among the 
existing $1D$ systems, one possibility is to use quantum wires 
which can be coupled easily to the SAW. However, since the densities for 
which adiabatic transport is most pronounced correspond to a few electrons 
per SAW spatial period (realistically, ca. a few microns), one would need 
wires with low electron $1D$ density of around $10^4{\rm cm}^{-1}$. 
The densities currently available in such systems
are at least an order of magnitude higher~\cite{q-wires}.

\begin{figure}
\centerline{\psfig{file=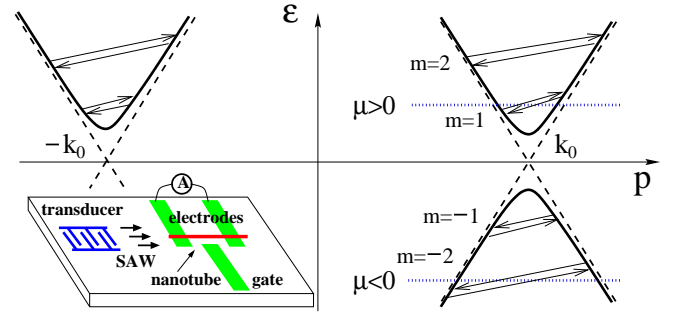,width=3.4in}}
\vspace{0.5cm}
\caption[]{The low energy spectrum of a metallic Carbon nanotube (broken line)
acquires a minigap (solid line) in the presence of a symmetry breaking 
perturbation. The backscattering transitions induced by the SAW 
potential are shown. {\it Inset:\ } proposed experimental arrangement 
consisting of a nanotube suspended between contacts, with a gate on
the side, and a SAW source.}
\label{fig:transitions}
\end{figure}

In this letter we argue that a surface acoustic wave (SAW) coupled to a
semi-metallic carbon nanotube presents an ideal system in which quantized
transport can be realized. The experimental arrangement is illustrated in
Fig~\ref{fig:transitions}. A nanotube is placed between two metallic 
contacts on the surface of a piezoelectric crystal, with a gate electrode 
nearby to allow adjustment of the Fermi level in the tube. In a 
piezoelectric substrate the SAW is accompanied by a wave of electrostatic 
potential that can have an amplitude up to few Volts\cite{SAW-book}.
The potential decays both into the free space 
and into the substrate 
to a depth comparable to the wavelength $\lambda_{\rm saw}$. 
We assume that the tube is suspended at a height $\ll \lambda_{\rm saw}$ 
above the substrate, so that there is no direct mechanical coupling  
and only the free space component of a SAW potential matters.
When a SAW is launched from a transducer (such as an inter-digitated electrode 
array) it's electric field penetrates the tube and electron diffraction 
on the sliding SAW potential results in miniband formation. By 
positioning the Fermi level within the energy gap, the conditions for 
current quantization are fulfilled.

High electron velocity in nanotubes, $v\approx 8\times 10^7{\rm cm/s}$, 
makes it possible to create large minigaps. 
The expected minigap size can be estimated in view of the observation~\cite{Tans,Bockrath'97}
of resonant states formed as standing waves in a finite sample. 
The resonances were found to be spaced by $hv/L$,
where $L$ the sample length. This gives $E_{\rm gap}=0.6 {\rm 
meV}$ for $L=3\mu{\rm m}$, which implies that a periodic SAW potential
with $\lambda_{\rm saw}$ of several microns is sufficient to form large minigaps.
A period of the SAW-induced 
grating down to $200-300{\rm nm}$ can be realized, so that minigaps as large as 
$10{\rm meV}$ are expected. For comparison, the same periodic 
perturbation acting on a GaAs $1D$ channel will induce minigaps of an 
order of magnitude smaller because of a smaller $v_F\approx 10^7{\rm cm/s}$.
A further advantage of metallic nanotube system results from its semi-metallic 
spectrum in which 
two pairs of oppositely moving spin degenerate states 
intersect exactly at the Fermi
level (at half-filling). Thus despite the fact that the SAW wavelength is always much larger 
than the lattice constant, principal minigaps will open close to the Fermi 
level (Fig.~\ref{fig:transitions}) and moreover a minute doping or gating 
is sufficient to align the chemical potential with one of the minigaps.

Electron states in semi-metallic nanotubes are described by the $1D$ Dirac
equation rather than the Schrodinger equation. We shall show that a
selection rule protects the integrity of the Dirac band structure against
backscattering due to a potential perturbation. Therefore in the
arrangement shown in Fig.~\ref{fig:transitions}, the SAW will not couple 
to the electrons at all! In order to realize adiabatic charge transport, 
backscattering must be restored by applying an external perturbation 
that lowers the symmetry of the Dirac system (by mixing left and right 
states). This can be achieved by applying a magnetic 
field~\cite{B-paral} along the nanotube axis. Also, in a number of 
nominally metallic nanotubes such as the so-called ``zig-zag'' nanotubes, 
a matrix element mixing the left and right states appears~\cite{zigzag} 
due to the curvature of the $2D$ Carbon sheet rolled into a tube. Both 
effects open a minigap at the band center, as confirmed 
experimentally~\cite{exp-B-parallel,exp-curvature}.
Below we consider electron spectrum in this system in the presence of 
a SAW potential, in the free electron model~\cite{Dresselhaus} and in the 
Luttinger liquid theory~\cite{KaneBalentsFisher,NT-luttinger}. We demonstrate that 
the electron interaction enhances the minigaps and, because of 
the $1D$ screening, scales up required SAW amplitude. 


The long-range electron interaction in the spin- and valley-degenerate 
modes is symmetric with respect to the four `flavors'. In the Luttinger 
liquid theory of nanotubes~\cite{KaneBalentsFisher,NT-luttinger} this 
interaction is described by the forward scattering amplitude $V(q)$ with 
a form that depends on the electrostatic environment. In the absence of 
screening, $V_0(q)
= e^2 \ln[(qd)^{-2}\!+\!1]$, where $d$ is the nanotube diameter. 
The substrate dielectric constant $\epsilon$ changes $V_0(q)$ to $V(q) = 
2V_0(q)/(\epsilon + 1)$. Since the ratio $N=d/a$ of the diameter $d$ to the 
inter-atomic distance $a$ is large (ca. $10$), backscattering and 
the Umklapp interactions are small scaling as $1/N$~\cite{KaneBalentsFisher}. 
Moreover, the Umklapp vertex also happens to be small 
numerically~\cite{YoshiokaOdintsov}.

Therefore, taking into account the presence of a symmetry breaking 
perturbation $\Delta$, and neglecting both backscattering and Umklapp 
processes, the low energy states of the nanotube system (in the 
vicinity of the band-crossing $p\!=\!\pm k_0$) are described by the Dirac 
Hamiltonian 
\bea\label{H-coulomb}
{\cal H}=\int\! dx\! 
\sum_{{\alpha}=1}^4\bar\psi_{\alpha}\left[-\hbar v\sigma_2\partial_x
+\Delta\right]\psi_{\alpha} 
+\frac{1}{2}\sum_q\widehat\rho_qV(q)\widehat\rho_{-q}
\eea
where $\bar\psi=\psi^\dagger\sigma_1$. 
Here Pauli matrices $\sigma_{1,2}$ operate in the two-component 
Dirac operator space $\psi_{\alpha}=(\psi_{\rm r},\psi_{\rm l})_{\alpha}$ with pseudospin 
components corresponding to the right and left moving states, and $\widehat\rho(x)
=\sum_{\alpha}\psi^\dagger_{\alpha}(x)\psi_{\alpha}(x)$ is charge density operator.
The second term in (\ref{H-coulomb}) 
describes the left/right mixing and yields a gap in the spectrum.
Different mixing mechanisms lead to different values of $\Delta$. For 
example, a parallel magnetic field~\cite{B-paral} produces $\Delta=\hbar 
v\phi/R$, where $R$ is the nanotube radius and $\phi=\Phi/\Phi_0$ is the 
magnetic flux through nanotube cross-section (measured in units of the flux 
quantum $\Phi_0=hc/e$). 

%
%

The harmonically varying electrostatic 
potential of the SAW decays exponentially in the direction normal to the 
surface: $A e^{-kz}\sin k(x-ut)$, where $u$ is SAW velocity. 
Since the wavelength 
$\lambda_{\sc saw}=2\pi/k$ is much larger than the tube diameter $2R$, 
one can ignore the potential variation $e^{-kz}$ over the tube 
cross-section. For example, for $\lambda_{\sc saw}=1\,\mu{\rm m}$ and 
$R=1\,{\rm nm}$, the potential variation is less than $1\%$. Therefore, it 
is sufficient to take into account just the parallel component of the 
SAW electric field. Since the SAW velocity is small, $u\ll v$, the
spectrum can, therefore, be obtained within a stationary approximation.

To simplify our analysis, let us first consider the non-interacting system.
In the stationary approximation, the single-particle 
spectrum of each degenerate `flavor' can be obtained from the perturbed 
1D Dirac system,
\begin{equation}
\epsilon\,\psi(x)=\lp -i\hbar v\partial_x \sigma_3+\Delta \sigma_1 + 
A \sin kx \rp \, \psi(x)\ .
\label{eq:dirac}
\end{equation}
Here the selection rule described earlier is manifest: for $\Delta=0$, 
Eq.~(\ref{eq:dirac}) separates into two independent equations for right and 
left moving particles. The SAW affects only the phase of the wavefunction. 
For $\Delta\ne 0$, the backscattering effect of the SAW potential is 
restored, and minigaps are induced in the spectrum. 

\begin{figure}
\centerline{\psfig{file=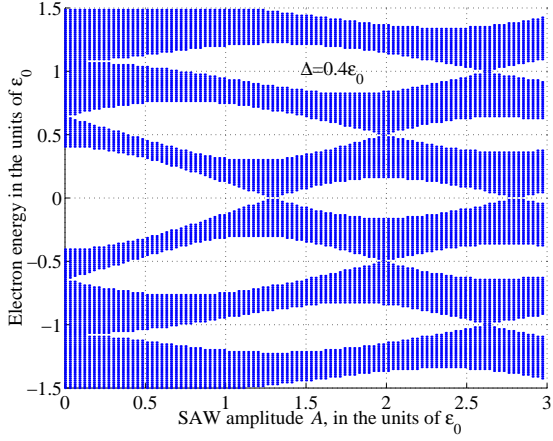,width=2.9in}}
\vspace{0.5cm}
\caption[]{Electron energy spectrum of Eq. (\ref{eq:dirac}) 
{\it vs.} the SAW field strength $A$, scaled by $\epsilon_0=
\hbar kv$.
Minigaps oscillate as a function of $A$, in agreement with the 
perturbation theory (\ref{eq:Delta-m}), 
vanishing at values close 
to the roots of Bessel functions.}
\label{fig:minibands}
\end{figure}

To explore the miniband structure of the Dirac system it is convenient to 
implement a gauge transformation, $\psi(x) = e^{\frac{i}2\sigma_3 \lambda
\cos kx} \psi'(x)$, where $\lambda=2A/\hbar kv$, and
\begin{equation}
\epsilon\,\psi'(x) =\lp -i \hbar v \sigma_3 \partial_{x}+\Delta \, 
e^{ -i \lambda \sigma_3 \cos kx} \, \sigma_1 \rp \psi'(x) 
\label{eq:ODE-main}
\end{equation}
The periodic system is characterized by Bloch states $\psi_p(x)=u_p(x)
e^{ipx}$ with quasimomentum $p$ taking values in the Brillouin zone defined 
by the SAW period, $-k/2<p<k/2$. The corresponding energy spectrum can 
easily be obtained numerically (Fig.\ref{fig:minibands}), by integrating 
the system of first-order differential equations (\ref{eq:ODE-main}) over 
the SAW spatial period $0<x<2\pi/k$. The spectrum has an electron-hole 
symmetry, $\epsilon\to -\epsilon$, characteristic of a Dirac system.  

The problem (\ref{eq:ODE-main}) can also be solved analytically 
for $\Delta \ll \hbar kv $, by treating the 
second term of Eq.~(\ref{eq:ODE-main}) as a perturbation~\cite{Yakovenko'01}. 
Separated into Fourier components,
\be
e^{-i\lambda\cos kx}=\sum_{m=-\infty}^{\infty}(-i)^mJ_m(\lambda) e^{-imkx},
\label{eq:Fourier}
\ee
where $J_m(\lambda)$ are Bessel functions, each harmonic of the perturbation
(\ref{eq:Fourier}) mixes right and left modes with $p-p'=mk$. When these 
states are in resonance (i.e., when $p=-p'=mk/2$, $\epsilon'_m=m kv/2$), 
the spectrum can be found by standard two-wave matching. This gives energy 
gaps 
\be
\label{eq:Delta-m}
\Delta_m= 2\Delta\, |J_m(2A/\hbar kv)|
\ee
which are oscillatory functions of the SAW amplitude $A$, with zeros
at the nodes of Bessel functions. 
In particular, for $A\ll\hbar kv$, $\Delta_m\simeq 2\Delta 
(A/\hbar kv)^{|m|}/m!$. 
%

%
Electrons in the half-filled (undoped) system represent a solid state 
analogue of the Dirac vacuum: under the SAW perturbation, the many-body
state carries neither charge nor current. For a weak SAW potential this 
  follows from
adiabatic continuity:
quantized transport takes place when the chemical potential 
$\mu$ falls in one of the minigaps. 
The value of the quantized current will remain the same~\cite{Thouless} within a whole 
range of values of $\mu$ and $A$ that stay within a gap.
Since the spectral gap at the band center is
adiabatically connected to the minigap at $A=0$ (induced by the
symmetry breaking perturbation $\Delta$), it is evident 
that at half-filling the current is zero. Similarly, for $m$ fully occupied minibands 
taking into account the four-fold valley and 
spin degeneracy, the electron density (counted from that at $\mu=0$) 
is $\delta n=4m\,k/2\pi$. This results 
in a current $j=eu\,\delta n$. Identifying $uk/2\pi$ with the SAW frequency 
$f$, we obtain quantized current $j=4m\,ef$. The dependence of the
energy gaps on $A$, shown in Fig.~\ref{fig:minibands}, describes the 
width of the plateaus of quantized current.

To complete our analysis it remains only to explore the integrity of the 
current quantization in the presence of electron interactions. To undertake 
this program it is convenient to first bosonize the 
Hamiltonian~(\ref{H-coulomb}) 
setting $\psi_j(x)\propto \exp\lp i\sqrt{\pi}\phi_j(x)\rp$. Introducing the
linear combination of bosonic fields
\begin{eqnarray*}
\lp \matrix{\Phi_0\cr \Phi_1\cr \Phi_2\cr \Phi_3}\rp
=\frac12
\lp \matrix{1 & 1 & 1 & 1\cr 1 & -1 & 1 & -1\cr
1 & -1 & -1 & 1\cr 1 & 1 & -1 & -1 }\rp
\lp \matrix{\phi_1\cr \phi_2\cr \phi_3\cr \phi_4}\rp ,
\end{eqnarray*}
the part of the Hamiltonian (\ref{H-coulomb}) without the mass term $\Delta\bar\psi\psi$ is 
diagonalized. Setting $V_{\rm ext}(x)=V_{\rm g} + A\sin kx$, where 
$V_{\rm g}$ represents the external gate potential, and 
$K(q)=1+4V_0(q)/\pi\hbar v$, the corresponding Lagrangian
  \bea\label{L0-bos}
&&{\cal L}_0 = \frac{1}{2}\sum_q\lb \p_{\tau}\Phi_0(q) \p_{\tau}\Phi_0(-q)
+ K(q) q^2\Phi_0(q)\Phi_0(-q)\rb \nonumber\\
&&\quad + \int \rd x \lb \frac{1}{2}\sum_{a = 1}^3(\p_{\mu}\Phi_a)^2 + 
\frac2{\sqrt{\pi}\hbar v}V_{\rm ext}(x)\Phi_0(x)\rb
  \eea
describes the dynamics of one charged and three neutral modes. Restoring the
mass term perturbation, the total Lagrangian is given by ${\cal L}=
{\cal L}_0+{\cal L}_\Delta$, where
  \be\label{Lext-bos}
{\cal L}_\Delta=-2\Delta\,\int \rd x \sum_{{\alpha} = 1}^4 \cos(\sqrt{4\pi}\phi_{\alpha})
\ .
  \ee
Applied to ${\cal L}$, the conventional RG approach demonstrates 
that the perturbation ${\cal L}_\Delta$ is relevant and grows. Depending on 
the density, controlled by $V_{\rm g}$, the resulting state can be gapped 
with a finite correlation length, or gapless.

Let us first focus on the influence of electron interactions on the energy 
gap at the band center considering the system at half-filling and in the 
absence of the SAW (i.e. $V_{\rm ext}=0$). Technically, this involves 
estimating the energy of a soliton field configuration $\phi_j$ (with any
flavor $j$). A variational analysis which takes into account the
renormalization due to the three neutral modes, obtains
%
%
  \be
E_{\rm gap} \approx K^{1/2}
E_0^{1/5}\Delta^{4/5}, 
\ {\rm where}\ E_0=\frac{\hbar v }{d},
  \ee
substantially {\em larger} than the non-interacting result, $\Delta$. 

Similarly, the SAW-induced minigaps (\ref{eq:Delta-m}) are also enhanced by 
interaction. Considering the regime $\Delta\ll \hbar kv$,
this enhancement is most straightforwardly demonstrated by mapping the 
SAW-induced gap onto the gap at the band center. This is achieved by a 
variable shift, 
  \be\label{phi-shift}
\phi_j\to\phi_j-\lp\sqrt{\pi}\hbar v\rp^{-1}\int^x_0
\widehat K^{-1} V_{\rm ext}(x')dx', 
  \ee
eliminating the term linear in $\Phi_0$ from Eq.~(\ref{L0-bos}). 
[The operator in (\ref{phi-shift}) is diagonal in Fourier representation, 
$\widehat K=K(q)$.]
At the same time the mass term~(\ref{Lext-bos}) is transformed as
  \be\label{Lext-bos-shifted}
{\cal L}_\Delta=-\Delta \int \rd x \sum_{f=1}^4 e^{i\lp \sqrt{4\pi}\phi_i
+\widetilde\lambda\cos kx-2\widetilde V_{\rm g} x \rp} + {\rm c.c.}
 ,
  \ee
where $\widetilde\lambda=2A/K\hbar kv$, $\widetilde V_{\rm g}=
V_{\rm g}/K\hbar v$, and $K=1+\frac8{hv}V(k)$. (Indeed, the shift (\ref{phi-shift}) is nothing but 
the bosonization representation of the gauge transformation used to solve 
the free fermion Dirac equation (\ref{eq:dirac}).) Now, by analogy with the 
treatment of Eq.~(\ref{eq:ODE-main}) above, one can expand ${\cal L}_\Delta$
in Fourier components $m$. The density corresponding to $m$ filled minibands 
can be chosen by setting $\widetilde V_{\rm g}=mk/2$. In this case, all terms 
in the Fourier series (\ref{eq:Fourier}) with $m\ne 2 \widetilde V_{\rm g}/k$
give rise to expressions with oscillatory spatial dependence. Discarding these
non-resonant terms one arrives at an expression of the form (\ref{Lext-bos})
with $\Delta$ replaced by $\Delta J_m(\widetilde\lambda)$. Being now formally
equivalent to the problem at half-filling considered above, we deduce that 
the interaction brings about a renormalization of the minigap such that
  \be
E_{\rm gap}^{(m)} \approx K^{1/2} E_0^{1/5}\big|\Delta J_m\lp 2A/K\hbar kv
\rp\big|^{4/5}
\ . 
  \ee
Several features of this result are worth noting: the general form of the 
energy gap dependence on the SAW amplitude, with nodes at the roots of Bessel 
functions, is {\em unaffected} by electron interaction. The magnitude of the 
minigap is {\em enhanced} by ca. $K^{1/2}\lp E_0/\Delta \rp^{1/5}$ as compared to the 
non-interacting case. The rescaling of the SAW amplitude $A$ and of 
$V_{\rm g}$ by $K$,
manifest in 
Eq.~(\ref{Lext-bos-shifted}), describes the effect of screening due to the
1D electron system. For the substrate dielectric constant $\epsilon=12$ 
(which corresponds to GaAs) we estimate the screening factor as $K\simeq 15$. 

To complete our discussion, let us comment on the feasibility of the 
experiment (Fig.~\ref{fig:transitions} inset). 
Maximal values of the 
SAW induced minigaps in Fig.~\ref{fig:minibands} are close to $\Delta$, one 
half of the value of the central gap. If 
a longitudinal magnetic field is used to open the 
central gap, then for a single-walled nanotube with a diameter 
$1.6\,{\rm nm}$ (such as that grown by Ref.~\cite{grown-NT}), and a 
field $B=16\,{\rm T}$, one finds $\Delta\simeq 5\,{\rm meV}$. Applied
to the spectra in Fig.~\ref{fig:minibands} where $\Delta=0.4\epsilon_0$ 
(i.e. $\epsilon_0=12\,{\rm meV}$), it corresponds to a SAW wavelength of 
$\lambda_{\sc saw}\simeq 0.25\,\mu{\rm m}$, frequency $f=13\,{\rm GHz}$,
and quantized current of around $8\,{\rm nA}$.
In order to reach a maximum value of the principal SAW induced minigap
shown in Fig.~\ref{fig:minibands} the SAW potential should be around 
$A=10\,{\rm meV}$. This value obtained in the single electron approximation 
should be corrected by the factor $K\sim 15$ to account for the screening 
in the 1D system. Thus a SAW potential of around several hundred meV may 
be required. These values do not present a problem even when 
a weak piezoelectric such as GaAs is used.\cite{Talyanskii'96} 
Moreover, for the experiments with nanotubes one can use a much stronger 
piezoelectric such as LiNbO${}_3$ as a substrate, which will make SAW potential
in the eV range available. A strong piezoelectricity of the substrate
will also facilitate generation of the high frequency SAW required for the 
proposed experiment (in LiNbO${}_3$ the SAW frequencies of ca. $17\,{\rm GHz}$ 
have been reported~\cite{Yamanouchi}). One could also use the 
``zig-zag'' nanotubes in which the central gap opens~\cite{zigzag} 
due to the curvature of the carbon sheet. In this case the gap 
is predicted~\cite{zigzag} to be in a range up to $20\,{\rm meV}$ 
for a tube diameter of $1.6\,{\rm nm}$, and the magnetic field is not necessary. 
Thus a SAW induced minigap
as large as $10\,{\rm meV}$ could be obtained in this case. 


To summarize, we have considered a metallic carbon nanotube in the field 
of a slowly moving periodic potential. If the nanotube is subjected to a 
further perturbation that mixes right and left moving states, the coupling
between the electrons and the SAW potential acts as a charge pump
conveying electrons along the tube. An estimate of the miniband spectrum 
induced by electron diffraction on the sliding potential revealed that 
minigaps of ca. $10\,{\rm meV}$ are viable. We therefore conclude
that the carbon nanotube combined with the SAW provides a promising system
in which quantized adiabatic charge transport can be observed. 
As demonstrated above, 
the energy gaps, that can be detected experimentally through quantization 
plateaux widths, are sensitive to the character of electron interactions. 
Thus, quantized transport in this strongly interacting system can be viewed as 
a novel probe of the Luttinger liquid physics. 

L.L. and D.N. acknowledge with pleasure the hospitality of the TCM Group 
at the Cavendish Laboratory. This work was supported by the MRSEC Program 
of the NSF 
under Grant No. DMR 98-08941.


\end{document}